# IVOA Compliant Services for the MACHO Data Archive

**Jonathan G. Smillie**[1]

**For the MACHO Collaboration**

## Abstract

The MACHO Project generated two-colour photometric lightcurves for 73 million stars in the LMC, SMC, and the Galactic bulge during its 8 years of observing. This photometry, along with all images from the over 100 thousand observations from which it was derived, and an associated catalogue of 21 thousand LMC variable stars, is now available via web-services which comply with standards defined by the International Virtual Observatory Alliance (IVOA).

## Introduction

The MACHO Project (Alcock et al. 2000) had as its principal goal the detection of gravitational microlensing in the LMC, the SMC and the Galactic bulge. The 50" Great Melbourne Telescope at Mt. Stromlo Observatory in Australia collected the observations via a CCD mosaic camera comprising 8 2k x 2k chips generating images in two passbands simultaneously. In excess of 100 thousand observations were taken between 1992 and 2000. Approximately 8TB of raw images resulted, from which photometric lightcurves were constructed for approximately 73 million stars. Subsequently, a catalogue of 21 thousand variable stars in the LMC was distilled from these lightcurves.

---

[1] Supercomputer Facility, Australian National University, Canberra, ACT 2600, Australia. Email: Jon.Smillie@anu.edu.au

Since its formation in June 2002, the International Virtual Observatory Alliance (IVOA) has published a range of data and service standards intended to facilitate the construction of an integrated international virtual observatory. These standards define services such as ConeSearch and Simple Image Access (SIA), which enable a data set to be searched based on user-defined sky regions, and VOTable, which is a data exchange format intended to facilitate the interoperation of disparate astronomical data sets and compatible tools.

## IVOA Compliant MACHO Services

Recent work at the Australian National University Supercomputer Facility (ANUSF) has resulted in the implementation of a range of IVOA compliant services which enable users to search and retrieve data from the full MACHO data set. The complete MACHO image metadata can be searched via both ConeSearch and SIA protocols, and the corresponding images can be downloaded directly to the user's desktop. Similarly, the macho star catalogue, including entries for 21 thousand LMC variable stars, can be searched via ConeSearch, and any of the lightcurves for the 73 million stars in the catalogue can be downloaded.

Services defined by IVOA protocols are in general intended to be accessed via compatible software tools, rather than by direct browser interaction. However simple form-based web-pages have also been implemented which allow users to directly query the MACHO services. Alternatively, a range of VO-compliant software tools are now available, such as TOPCAT (TOPCAT), which are able to directly access IVOA compliant data services, such as those discussed here.

## MACHO Service URLs

The IVOA compliant MACHO services can be accessed via the server at http://macho.anu.edu.au. Four services are available from this page, namely observation metadata search via ConeSearch and SIA protocols (http://macho.anu.edu.au/image), observation image download (http://macho.anu.edu.au/image_download), star catalogue search, including the variable star catalogue, via ConeSearch (http://macho.anu.edu.au/star), and photometry search and download via ConeSearch (http://macho.anu.edu.au/photometry). These URLs all access simple search forms suitable for direct user interaction. Note these form based interfaces are NOT compliant with any specific standard, and are provided merely as a user convenience.

The underlying IVOA compliant services are accessed directly via the following URLs. The image metadata catalogue can be searched via ConeSearch at http://macho.anu.edu.au/image/conesearch, or SIA at http://macho.anu.edu.au/image/sia. The VOTable results returned by these two services include URLs which enable the corresponding images to be downloaded. ConeSearch of the star catalogue (including variable stars) can be performed at http://macho.anu.edu.au/star/conesearch. ConeSearch of the photometry lightcurve data can be performed at http://macho.anu.edu.au/photometry/conesearch. It is these URLs to which compliant software tools should be directed, rather than the URLs accessing search forms described above. Interaction with these services is as per the associated protocol definition.

## Available Documentation

Some documentation for these services, and their usage, has been provided. This includes a user guide (https://dc2.apac.edu.au/trac/macho/browser/trunk/macho/doc/user_guide.txt) and a guide regarding how to perform some typical operations using these services (https://dc2.apac.edu.au/trac/macho/browser/trunk/macho/doc/how_to.txt).

## Service Compliance

At the time of writing, the ConeSearch services described here comply with IVOA ConeSearch Version 1.03 (http://www.ivoa.net/Documents/REC/DAL/ConeSearch-20080222.html), and the SIA service complies with IVOA SIA Version 1.0 (http://www.ivoa.net/Documents/SIA/20091008/PR-SIA-1.0-20091008.html). It should be noted that whilst the referenced ConeSearch standard is a formal IVOA Recommendation, the referenced SIA standard is at the time of writing only a Proposed Recommendation, which is in its Request-For-Comment phase, and may be revised at any time.

All services return VOTable Version 1.1 XML documents (http://www.ivoa.net/Documents/REC/VOTable/VOTable-20040811.html).

All services have been validated with the US National Virtual Observatory (NVO) validation services at http://nvo.ncsa.uiuc.edu/dalvalidate/csvalidate.html (ConeSearch) and http://nvo.ncsa.uiuc.edu/dalvalidate/siavalidate.html (SIA), and found to comply fully with the above protocol standards. They have also been validated by the NVO automatic validation service, whose results are available at http://heasarc.gsfc.nasa.gov/vo/validation/. Searching these results by the centre code "ANU" will yield the validation results for the services discussed here.

## Service Registration

All services have been registered with the NVO service registry (http://nvo.stsci.edu/vor10/), and will thus be available to all tools which utilize this registry. A search on this registry using the terms "macho anusf" will yield the registry entries for the services described here.

## Compatibility with Earlier Services

The services described here are intended to functionally replace the services described in Allsman et al. (Allsman et al. 2001) which have been in constant use now for approximately eight years.

## Acknowledgements

The authors would like to acknowledge the contributions, in terms of advice, suggestions, comments, and the example set by the original MACHO web services, of R. A. Allsman. We would also like to acknowledge our colleagues at ANUSF for their technical support and assistance, without which this work would have taken a lot longer, and may have resulted in something quite different, in particular S. Hungerford and J. Ozolins, with whom various discussions proved enormously useful.

This work was supported, at various times, by the following grants: ARC LIEF LE0561221; APAC Data Intensive Projects 2005-7; ANUSF merit round data grant d01.

## References

Alcock, C., et al. 2000, ApJ 542, 281

Allsman, R.A., et al. 2001, http://arxiv.org/abs/astro-ph/0108444v1

ANUSF: http://anusf.anu.edu.au/

IVOA: http://www.ivoa.net/

NVO: http://www.us-vo.org/

TOPCAT: http://www.star.bris.ac.uk/~mbt/topcat/